\documentclass[pra,singlecolumn,,amsmath,amssymb,floatfix]{revtex4}
\usepackage{graphicx}
\usepackage{bm}
\usepackage{amsmath,amssymb}

\def \be{\begin{equation}}
\def \ee{\end{equation}}
\def \ba{\begin{array}}
\def \ea{\end{array}}
\def \beq{\begin{eqnarray}}
\def \eeq{\end{eqnarray}}

\def \bed{\begin{displaymath}}
\def \eed{\end{displaymath}}

\begin{document}

\title{Universal Dynamics Near Quantum Critical Points}

\author{Vladimir Gritsev$^1$ and Anatoli Polkovnikov$^2$}
\affiliation{$^1$Physics Department, University of Fribourg, Chemin du Musee 3, 1700
Fribourg, Switzerland\\
$^2$Department of Physics, Boston University, 590 Commonwealth Ave., Boston, MA 02215, USA}

\begin{abstract}

We give an overview of the scaling of density of quasi-particles and excess energy (heat) for nearly adiabatic dynamics near quantum critical points (QCPs). In particular we discuss both sudden quenches of small amplitude and slow sweeps across the QCP. We show close connection between universal scaling of these quantities with the scaling behavior of the fidelity susceptibility and its generalizations. In particular we argue that the Kibble-Zurek scaling can be easily understood using this concept. We discuss how these scalings can be derived within the adiabatic perturbation theory and how using this approach slow and fast quenches can be treated within the same framework. We also describe modifications of these scalings for finite temperature quenches and emphasize the important role of statistics of low-energy excitations. In the end we mention some connections between adiabatic dynamics near critical points with dynamics associated with space-time singularities in the metrics, which naturally emerges in such areas as cosmology and string theory.

\end{abstract}
\maketitle

Continuous quantum phase transitions (QPTs) have been a subject of
intense theoretical research in recent decades (see e.g.
Refs.~\cite{sachdev_book, sondhi1997rmp,  vojta2003progphys} for
overview). Unlike usual phase transitions driven by temperature,
QPTs are driven entirely by quantum fluctuations. They are believed
to occur in many situations as described later in this book. Quite
recently a second order QPT was observed in a cold atom system of
interacting bosons in an optical lattice. There a system of
interacting bosons was driven in real time from the superfluid to
the insulating phase~\cite{greiner2002nature}, confirming an earlier
theoretical prediction~\cite{fisher1989}. Up till now, Ref.~\cite{greiner2002nature} has provided probably the cleanest
experimental confirmation of a QPT. The unifying property of all
continuous (second order) phase transitions is the emergent
universality and scale invariance of the long-distance low energy
properties of the system near the quantum critical point
(QCP)~\cite{sachdev_book}. This universality implies that low-energy
properties of the system can be described by very few parameters,
like the correlation length or the gap, which typically have
power-law scaling with the tuning parameter characterized by
critical exponents. These exponents are not sensitive to the
microscopic details of the Hamiltonian describing the system, but
rather depend only on the so-called universality class to which a given
phase transition belongs~\cite{sachdev_book}.

Recent experimental progress in preparing and manipulating out-of-equilibrium nearly isolated systems has stimulated intense theoretical research on quantum dynamics in closed systems. In particular, such issues as sudden quantum quenches in low dimensional systems~\cite{cardy_quench1, cardy_quench2}, adiabatic
dynamics near QCPs~\cite{ap_adiabatic, zurek_adiabatic}, and connection of dynamics and thermodynamics in
quantum systems~\cite{reimann, olshanii_nature} came to the forefront of theoretical research. A very natural question can be posed about
non-equilibrium behavior of systems near QPTs. Since in equilibrium
second order phase transitions are characterized by universality, one
can expect also universal behavior in driven systems in the vicinity
of a QCP. Such universality can be expected, for example, if the
system near the QCP is a subject to small amplitude, low frequency
modulation of an external field, which couples to the order
parameter. Indeed, in the linear response regime QCPs are typically
characterized by singular susceptibilities at low frequencies~\cite{sachdev_book,
vojta2003progphys}. One can expect that the universality persists
even beyond the linear response regime as long as the systems
remains close to the criticality. Another possible situation where
one can expect universal behavior is when the system is slowly
driven through a QCP. In this case, since the dynamics is nearly
adiabatic, one expects that low energy excitations will play the dominant role. Moreover, one can generally expect that non-adiabatic effects will be especially strong near singularities like QPTs, so that the dynamics will be dominated by the universal regime. This is indeed the case at least in sufficiently small dimensions~\cite{ap_adiabatic, zurek_adiabatic, pg_np}. The third situation where universality of dynamics can be expected is the response to fast small amplitude quenches (sudden changes in the tuning parameter) near the critical point. Analysis of sudden and nearly adiabatic dynamics near QCPs will be the central subject of this chapter.

Typically analysis of slow, nearly adiabatic dynamics, is complicated by the fact that the usual perturbative approaches fail. It has been recently realized that adiabatic perturbation theory can become an efficient tool for analyzing the behavior of various thermodynamic quantities like the density of quasi-particles and the non-adiabatic energy (heat) generated during the process~\cite{ap_adiabatic, claudia_kolk}. Although to leading order adiabatic perturbation theory often fails to give accurate quantitative analysis of various observables, it does give their correct scaling behavior with the rate of change of the external parameter.  The advantage of this method is that it allows one to effectively reduce a dynamical problem to the static one and use the developed machinery for equilibrium quantum phase transitions.  In this chapter we will demonstrate how this approach reproduces the correct scaling behavior of the response of the system both to sudden quenches of small amplitude and to slow nearly adiabatic sweeps through the critical point. We will also discuss the close connection between universal scaling of the density of quasi-particles and the scaling behavior of the fidelity susceptibility  near the QCP, the quantity characterizing the overlap of the ground state wave functions corresponding to different coupling constants. In particular, we will show that the Kibble-Zurek scaling~\cite{kz1,kz2} can be understood using this concept. We will illustrate some of the results using the transverse-field Ising model. In the end we will briefly discuss connections between adiabatic dynamics near critical points and dynamics associated with space-time singularities in the metrics, which naturally emerge in such areas as cosmology and string theory.

\section{Brief Review of the Scaling Theory for Second Order Phase Transitions}
\label{sec:review}

Continuous quantum  phase transitions in many respects are similar to classical second order phase transitions. The main difference is that the quantum transition from one phase to another is driven by quantum rather than by thermal fluctuations, arising from the zero-point motion. So QPTs can happen at zero temperature.  Examples of models where QPTs take place include quantum Ising and rotor models, the sine-Gordon model, various transitions from glassy to ordered phases in disordered systems and many others~\cite{sachdev_book}.  Many examples of QPTs will be discussed in consequent chapters in this book. Very often QPTs in $d$-dimensional systems can be mapped to thermal (classical) transitions in $d+z$ dimensions, where $z$ is the dynamical critical exponent~\cite{sachdev_book}. One of the most important properties of both quantum and classical second order transitions is the universality of the low-energy long-distance properties of the system. This universality implies that the details of underlying microscopic models are not important near critical points. Instead the properties of the system can be well characterized by the parameter describing its proximity to the critical point (tuning parameter) and by universal critical exponents describing the singular behavior of various quantities with this parameter. We note that even though QPTs strictly speaking occur only at zero temperatures, the universality of the scaling governed by QCPs extends well into the finite temperature domain~\cite{sachdev_book}. Recently there has been considerable interest in unconventional phase transitions, the description of which requires deviations from the standard framework. Some of these unconventional transitions will be discussed in other chapters of this book. Here we will not consider them since their dynamics is not yet well understood.

A key quantity characterizing continuous phase transitions is the correlation length $\xi(\lambda)$, which defines the length scale separating the qualitatively different behavior of e.g. spatial correlation functions of the fluctuations of the order parameter. This length scale diverges with the tuning parameter $\lambda$ as
\begin{equation}
\xi(\lambda)\sim 1/|\lambda-\lambda_c|^\nu,
\label{corr_length}
\end{equation}
where $\nu$ is a critical exponent. Divergence of the length scale is accompanied by divergence of associated time scales. For classical phase transitions the corresponding time scales are associated with relaxational dynamics. In quantum systems by the uncertainty principle a divergent time scale is also characterized by a vanishing energy scale $\Delta(\lambda)$. The latter is usually associated with a crossover in the quasi-particle spectrum. E.g. it can represent a gap, or an energy scale where the dynamical exponent changes. Scaling of $\Delta$ defines another dynamical critical exponent $z$:
\begin{equation}
\Delta(\lambda)\sim 1/\xi(\lambda)^z\sim |\lambda-\lambda_c|^{z\nu}.
\label{eqn:polkovnikov2}
\end{equation}
There are many other critical exponents, however, these two will play the most important role in our discussion. Together with critical exponents, one can introduce scaling functions which describe long-distance low-energy properties of the system. For example, if we are talking about a QPT in a spin system then near the critical point we expect the equal-time correlation between the spin $s$ at different positions to scale as
\begin{equation}
\langle s(x) s(0)\rangle\sim |x|^{-2\alpha} F(x/\xi).
\end{equation}
where $\alpha$ is the scaling dimension of the spin $s$ and $F$ is
some scaling function which approaches a constant when $x/\xi\ll 1$.  (Note that the divergence of the correlation functions at very small $x$ will be cut off by non-universal short-distance physics).

Other important quantities characterizing continuous phase transitions are susceptibilities, which describe the response of the system to external perturbations. For example, for a spin system the magnetic susceptibility describes the response of the magnetization to a small modulation of the magnetic field. From standard perturbation theory it is well known that susceptibilities are closely related to correlation functions~\cite{sachdev_book}. Near QCPs static susceptibilities usually have singular non-analytic behavior characterized by their own critical exponents (see e.g. Ref.~\cite{vojta2003progphys}). Recently, it was realized that a very useful measure to analyze quantum phase transitions is fidelity susceptibility (FS) $\chi_f$ or more generally a quantum geometric tensor~\cite{venuti}. As we will see, this equilibrium concept and its generalizations will be very important for us later when we analyze dynamics near QCPs. Formally in a d-dimensional system FS is defined as
\be
\chi_f(\lambda)={1\over L^d}\langle
\partial_\lambda\Psi_0(\lambda)|\partial_\lambda\Psi_0(\lambda)\rangle={1\over L^d}
\sum_{n\neq 0} \frac{|\langle \Psi_0(\lambda)|\partial_\lambda H|\Psi_n(\lambda)\rangle|^2}
{|E_n(\lambda)-E_0(\lambda)|^2},
\label{chif}
\ee
where $|\Psi_n(\lambda)\rangle$ denote instantaneous eigenstates of
the Hamiltonian $H(\lambda)$ and $E_n(\lambda)$ are the instantaneous energies. In this chapter we are only concerned with a non-degenerate ground state. FS appears in the leading order of expansion of the overlap of the ground state functions $|\langle
\Psi_0(\lambda+\delta\lambda)|\Psi_0(\lambda)|^2$ in the powers of $\delta\lambda$:
\be
|\langle \Psi_0(\lambda+\delta\lambda)|\Psi_0(\lambda)|^2\approx 1-\delta\lambda^2 L^d \chi_f(\lambda).
\ee
In the case when the system is translationally invariant and the operator $\partial_\lambda H$ is local, i.e., $\partial_\lambda H=\int d^d x v(x)$, where $x$ is the discrete or continuous coordinate, Eq.~(\ref{chif}) implies that the scaling dimension of $\chi_f$ is ${\rm dim}[\chi_f]=2\Delta_v-2 z-d$, where $\Delta_v$ is the scaling dimension of $v(x)$ (i.e. at the critical point $\langle v(x)v(x')\rangle \sim 1/|x-x'|^{2\Delta_v}$). This scaling dimension implies that the singular part of the FS near the critical point behaves as~\cite{venuti, gu}
\be
\chi_f(\lambda)\sim |\lambda-\lambda_c|^{\nu(2\Delta_v-2z-d)}.
\label{chif_scaling}
\ee
In finite size systems the singularity of $\chi_f$ at the critical point is cutoff by the system size so that
\be
\chi_f(\lambda_c)\sim L^{-2\Delta_v+2z+d}.
\label{chif_scaling1}
\ee
The crossover between the scalings (\ref{chif_scaling}) and (\ref{chif_scaling1}) occurs when the correlation length
(\ref{corr_length}) becomes comparable to the system size $L$. For relevant or marginal perturbations the product $\lambda v(x)$ should have the scaling dimension of the energy density: $d+z$. Because the scaling dimension of $\lambda$ is by definition $1/\nu$ we find that in this case $\Delta_v=d+z-1/\nu$~\cite{alet,quench_qcp}. Then Eqs.~(\ref{chif_scaling}) and (\ref{chif_scaling1}) for the singular part of $\chi_f$ reduce to
\be
\chi_f(\lambda)\sim \left\{\begin{array}{cc}
|\lambda-\lambda_c|^{d\nu-2} & \xi(\lambda)\ll L\\
L^{2/\nu-d} & \xi(\lambda)\gg L
\end{array}\right.
\label{chif_scaling2}
\ee
On top of the singular part there can be a non-singular part, which is generally non-sensitive to the proximity to the critical point. From Eq.~(\ref{chif_scaling2}) we see that the singular part of $\chi_f$ gives is dominant near the QCP when $d\nu<2$. The fidelity susceptibility can be expressed through the imaginary time connected correlation function~\cite{venuti}:
\be
\chi_f(\lambda)={1\over L^d}\int_0^\infty d\tau \tau \left( \langle \partial_\lambda H(\tau) \partial_\lambda H(0)\rangle-\langle\partial_\lambda H(0)\rangle^2\right),
\label{chif_corr}
\ee
where $\partial_\lambda H(\tau)=\exp[H\tau]\partial_\lambda H\exp[-H\tau]$. The above relation immediately follows from the Lehmann's representation. From Eq.~(\ref{chif_corr}) it is obvious that the singular scaling of $\chi_f$ near the QCP is related to the power-law long-distance asymptotics of the correlation function.

For the reasons which will become clear later, it is very convenient to generalize the fidelity susceptibility~(\ref{chif}) to the generalized adiabatic susceptibility of order $m$~\cite{quench_qcp}:
\be
\chi_m(\lambda)={1\over L^d}
\sum_{n\neq 0} \frac{|\langle \Psi_0(\lambda)|\partial_\lambda H|\Psi_n(\lambda)\rangle|^2}
{|E_n(\lambda)-E_0(\lambda)|^m}.
\label{chim}
\ee
For $m=2$ we recover the fidelity susceptibility: $\chi_2=\chi_f$. For $m=1$ the corresponding susceptibility describes the second order correction to the ground state energy. The scaling dimension of $\chi_m$ immediately follows from that of $\chi_f$: ${\rm dim}[\chi_m]={\rm dim}[\chi_f]-z(m-2)$. The generalized adiabatic susceptibility can also be expressed through the connected correlation function:
\be
\chi_m(\lambda)={1\over L^d (m-1)!}\int_0^\infty d\tau \tau^{m-1} \left( \langle \partial_\lambda H(\tau) \partial_\lambda H(0)\rangle-\langle\partial_\lambda H(0)\rangle^2\right).
\ee

{\em Example: transverse-field Ising model.} Let us illustrate some
of the generic properties mentioned above using the transverse-field
Ising model, which is perhaps one of the simplest models showing
quantum critical behavior. The Hamiltonian describing this
model is
\begin{equation}
H_I=-J\textstyle\sum_i \left(g\sigma^x_i+\sigma^z_i\sigma^z_{i+1}\right),
\label{h_ising}
\end{equation}
where $\sigma_i^x$ and $\sigma_i^z$ are the Pauli matrices,
commuting on different sites, $J$ is the energy scale for spin-spin coupling, and $Jg$ is the energy scale for the transverse field.
This model undergoes two QPTs at $g=\pm 1$. At small magnitude of
the transverse field $g$ the spins are predominantly magnetized
along the z-axis, while at large $|g|$ they are magnetized along the
direction of the magnetic field. In both cases there is a finite gap to the lowest energy excitations which
vanishes at the critical point~\cite{sachdev_book}. This model can
be solved exactly using the Jordan-Wigner transformation to map it to noninteracting fermions.

The Hamiltonian is diagonal in momentum space and can be rewritten as
\be
H_I=\sum_k \epsilon_k(\gamma_k^\dagger\gamma_k-1/2),
\label{h_ising1}
\ee
where $\epsilon_k=2\sqrt{1+g^2-2g\cos k}$ and $\gamma_{k},\gamma_{k}^{\dagger}$ are the fermionic operators. The
ground state of this Hamiltonian, which is the vacuum of $\gamma_k$, is written as a direct product:
\be
|\Psi_0\rangle=\bigotimes_{k>0} \left[\cos(\theta_k/2) +i\sin(\theta_k/2)c_k^\dagger c_{-k}^\dagger\right]|0\rangle,
\label{gs_tfim}
\ee
where $|0\rangle$ is the vacuum of original Jordan-Wigner fermions and $\tan\theta_k=\sin(k)/(\cos(k)-g)$.  Further details can be found in Sachdev's book~\cite{sachdev_book}.

It is convenient to introduce a tuning parameter $\lambda=g-1$ so that the transition occurs at $\lambda=\lambda_c\equiv 0$. Near the critical point the expressions for $\theta_k$ and dispersion simplify: $\tan\theta_k\approx -k/( k^2+\lambda)\approx -k/ \lambda$ and $\epsilon_k\approx 2\sqrt{\lambda^2+k^2}$. This scaling of energy immediately suggests that the critical exponents here are $z=\nu=1$. Indeed, according to our general discussion  (see Eq.~(\ref{eqn:polkovnikov2}))  the characteristic energy scale, which is obviously the gap in our case, scales as $\Delta=|\lambda|^{z\nu}$, which implies that $z\nu=1$. Also, the spectrum clearly has a crossover from a constant to a linear function of momentum at $k^\star\sim |\lambda|$, suggesting that there is a characteristic correlation length scales as $\xi\sim 1/k^\star\sim 1/|\lambda|$; this implies that $\nu=1$. Using the factorization property (\ref{gs_tfim}) it is easy to check that
\be
\chi_f={1\over L}\langle\partial_\lambda\Psi(0)|\partial_\lambda\Psi(0)
\rangle=\frac{1}{4 L}\sum_{k>0} \left(\frac{\partial \theta_k}{\partial\lambda}\right)^2\approx \frac{1}{4}\sum_{k>0} \frac{k^2}{(k^2+\lambda^2)^2}.
\label{chif_ising0}
\ee
In the thermodynamic limit for $\lambda\gg 1/L$ we can substitute the sum with the integral and find
\be
\chi_f(\lambda)=\int_0^\infty \frac{dk}{2\pi} {k^2\over (k^2+\lambda^2)^2}=\frac{1}{8\lambda}\,.
\label{chif_ising}
\ee
At the QCP and finite $L$ Eq.~(\ref{chif_ising0}) gives $\chi_f(0)\approx L^2/96$. Both asymptotics are consistent with the expected scaling~(\ref{chif_scaling2}).

\section{Scaling Analysis for Dynamics near Quantum Critical Points}
\label{sec:scaling}

Above we gave a brief overview of the universal aspects of the equilibrium ground state properties of quantum critical systems. Our next goal is to see how similar universality emerges in out-of-equilibrium situations where the parameter $\lambda$ is tuned in time through a QCP. We expect that the dynamics will be universal if
it is dominated by the low energy excitations generated near the critical point. In this chapter we will consider two specific situations, where such universality is found: (a) instantaneous quench of the small amplitude starting precisely at the critical point; and (b) slow quenches, where the tuning parameter changes gradually across the QCP. We will also allow for slow quenches to start (end) exactly at the QCP. These two situations correspond to two very generic scenarios. The first one is realized when the rate of change of the tuning parameter is fast compared to other relevant time scales. Though starting exactly at the critical point can require fine tuning, the scaling results will remain valid as long as the system is sufficiently close to the critical point both before and after the quench. The second situation is applicable to regimes in which the tuning parameter changes slowly in time. As we will show below these two scenarios can be actually described within the same unifying framework. In this section we will use qualitative arguments based on the scaling of the FS near the critical point. In the next section we will derive these scaling
relations more accurately using adiabatic perturbation theory.

We first start from instantaneous quenches. As we mentioned above, we assume the system is initially prepared in the ground state of some Hamiltonian $H_0$ corresponding precisely to the QCP.  We then suddenly apply perturbation $\lambda V$, where $\lambda$ is a small parameter and $V$ is some operator independent of $\lambda$. We will assume that $V$ is a relevant perturbation which drives the system away from the QCP. Let us discuss the expected scaling of various quantities using ordinary perturbation theory. According to general rules of quantum mechanics, the ground state wave function after the quench will be projected onto the basis of the new (quenched) Hamiltonian. Within ordinary perturbation theory the amplitude $\delta\psi_n$ to occupy the excited state $|n\rangle$ is
\be
\delta\psi_n\approx \lambda\frac{\langle n|V|0\rangle}{E_n-E_0},
\ee
where the matrix elements and the energies are calculated for the unperturbed Hamiltonian. It is clear that quantities which commute with the new Hamiltonian like the excitation probability or the energy should scale as a square power of $\lambda$. This follows from the fact that such quantities must be quadratic in $\delta\psi_n$. For example, the probability of exciting the system is \be
P_{\rm ex}=\sum_n^\prime |\delta\psi_n|^2=\lambda^2 \sum_n^\prime \frac{|\langle n|V|0\rangle|^2}{|E_n-E_0|^2}=\lambda^2 L^d \chi_f(0),
\label{G}
\ee
where the prime over the sum implies that the ground state is excluded from the summation. The excitation probability is meaningful only for quenches with the amplitude vanishing in the thermodynamic limit such that $\xi(\lambda)\gtrsim L$). For larger amplitude quenches $P_{\rm ex}\approx 1$. In these situations one needs other quantities to characterize the system's response to the quench. One of such quantities can be diagonal entropy~\cite{diag_entropy}: $S_d=-\sum |\psi_n|^2 \ln|\psi_n|^2$, where $|\psi_n|^2$ are the probabilities to occupy eigenstates of the quenched Hamiltonian. Clearly $S_d$ is a possible measure of the non-adiabatic response of the system, which is well defined for quenches of all amplitudes. Another possible measure characterizing the non-adiabatic response of the system is the energy, or more precisely the non-adiabatic part of the energy change due to transitions. This excess energy, or simply heat~\cite{ap_heat}, is the difference between the energy after the quench and the instantaneous ground state energy. Within the perturbation theory we find that the heat density $Q$ is
\be
Q(\lambda)=\frac{1}{L^d}\sum_n (E_n-E_0) |\delta\psi_n|^2=\frac{1}{L^d}\lambda^2 \sum_n^\prime \frac{|\langle n|V|0\rangle|^2}{E_n-E_0} \equiv \lambda^2\chi_E(0),
\label{Q}
\ee
where we introduced the energy susceptibility $\chi_E(\lambda)= \chi_1(\lambda)$. If we additionally assume that excitations in the system are carried by well-defined quasi-particles, i.e., that the energy eigenstates
$|n\rangle$ are also the eigenstates of the quasi-particle number operator (which is usually the case only in integrable models), then we can also compute the density of quasi-particles,
\be
n_{\rm ex}(\lambda)=\frac{1}{L^d}\sum_n N_n |\delta\psi_n|^2=\lambda^2\chi_n,
\label{Nex}
\ee
where $N_n$ is the total occupation of each eigenstate $n$
\be
\chi_n=\frac{1}{L^d}\sum_n^\prime N_n \frac{|\langle n|V|0\rangle|^2}{(E_n-E_0)^2}.
\ee
In most systems the number of quasi-particles is actually not conserved due to various collision processes. However, the quasi-particle number can be still a very useful quantity if we are interested in times shorter than the relaxation time, i.e., times in which collisions do not play any significant role. We see that in all these situations we expect $P_{\rm ex}(\lambda),\, Q(\lambda),$ and $n_{\rm ex}(\lambda)$ to scale quadratically in $\lambda$ to lowest order. The scaling for the d-entropy can acquire additional logarithmic corrections. We will not discuss it here for brevity since it is very similar to that of $n_{\rm ex}$~\cite{quench_qcp}.  We note that this quadratic scaling should be contrasted with the usual linear response relations where various quantities are proportional to the first power of $\lambda$, e.g.,
\be
\langle \psi(\lambda)| V|\psi(\lambda)\rangle\approx {\rm const}+2\lambda\sum_n {|\langle n|V|0\rangle|^2\over E_n-E_0}={\rm const}+2 L^d \lambda\chi_E.
\ee
To obtain linear dependence it is obviously crucial that the observable of interest does not commute with the Hamiltonian in the excited state. Such situations can also be analyzed. However, we will not consider them.

It is very important to remark that we are dealing with extended systems. As we saw in the simple example of the transverse-field Ising model the FS is extensive in the system size~(\ref{chif_ising}). This situation is generic if we consider global (spatially-uniform) perturbations. Strictly speaking this implies that the validity of the perturbation theory is restricted to very small perturbations, where $P_{\rm ex}(\lambda)\ll 1$. However, we know very well that perturbative approaches typically have much larger domains of applicability. It is usually important that only changes of intensive quantities like the energy per unit volume or the density of quasi-particles remain small.
Indeed, the probability to produce the next quasi-particle excitation is not affected much by the presence of other quasi-particles if those are very dilute. Thus we expect that Eqs.~(\ref{Q}) and (\ref{Nex}) have a much broader domain of applicability than Eq.~(\ref{G}) (see also Refs.~\cite{pg_np, quench_qcp} for additional discussion).

As we discussed earlier, near QCPs susceptibilities may become divergent (see e.g. Eq.~(\ref{chif_ising}) for the Ising model) or acquire some other type of singularity. Such divergences can invalidate the quadratic perturbative scaling
of various observables and these situations will be the next point of our discussion. Let us look into e.g. Eq.~(\ref{G}) for the probability of exciting the system. Near the critical point $\chi_f(\lambda)$ may diverge as $\lambda\to 0$ with the divergence being cutoff by the system size (see Eq..~\ref{chif_scaling2}). This results in the superextensive scaling of the transition probability:
\be
P_{\rm ex}(\lambda)\sim \lambda^2 L^{2/\nu}
\label{p_ex}
\ee
for $2/\nu>d$. It is clear that $P_{\rm ex}$ should saturate at unity when $\lambda$ becomes bigger than $\lambda^\ast\sim 1/L^{1/\nu}$, which is accidentally equivalent to the condition $\xi(\lambda)<L$ where the fidelity susceptibility changes its form (see Eq.~\ref{chif_scaling2}). Likewise for $2/\nu>d+z$ for the heat density we find
\be
Q(\lambda)\sim \lambda^2 L^{2/\nu-d-z}
\ee
for $\lambda<\lambda^\ast$. However, unlike the probability of excitations, we can extend the perturbative expression for the heat~(\ref{Q}) to the regime $\lambda>\lambda^\ast$ by simply evaluating the susceptibility $\chi_E$ at the final value of the quench parameter. More accurate derivation based on the adiabatic perturbation theory, which we discuss in the next section, shows that one needs to change $\chi_E(0)\to \int_0^\lambda \chi_E(\lambda')d\lambda'$ in Eq.~(\ref{Q}). However, the scaling of this integral is the same as the scaling of $\chi_E(\lambda)$. Then we find that for $\lambda>\lambda^\ast$:
\be
Q(\lambda)\sim |\lambda|^{(d+z)\nu}.
\label{q_scaling}
\ee
Finally if we additionally assume that dominant excitations are coming from isolated quasi-particles (which is often the case) we find that for $\lambda>\lambda^\ast$
\be
n_{\rm ex}\sim |\lambda|^{d\nu},
\label{n_scaling}
\ee
while for $\lambda<\lambda^\ast$ the scaling for $n_{\rm ex}$ is similar to that of $P_{\rm ex}$ given by Eq.~(\ref{p_ex}).

Of course, we expect these scalings to remain valid as long as the corresponding exponents do not exceed two. Otherwise, the low-energy singularities associated with the critical point become sub-leading (corresponding susceptibilities do not diverge) and the perturbative quadratic scaling is restored (though the singularities still appear in higher order derivatives with respect to $\lambda$).

There is a very simple argument, which reproduces the scalings (\ref{q_scaling}) and (\ref{n_scaling}). A quench of amplitude $\lambda$ gives us natural length and energy scales $\xi\sim 1/|\lambda|^\nu$ and $\Delta\sim
|\lambda|^{z\nu}$. We thus might expect that the quasi-particle excitations with energies larger than $\Delta$, or equivalently with momenta larger than $1/\xi$, will not be much affected by the quench and can be treated perturbatively, yielding the quadratic scaling. At the same time, for states with energies less than $\Delta$ the
quench will be effectively very strong, so that they will be excited with the probability of the order of unity. Thus the density of quasi-particles will scale as $1/\xi^d\sim \lambda^{d\nu}$ and we reproduce the scaling (\ref{n_scaling}). If $d\nu<2$ then this contribution to $n_{\rm ex}$ will be dominant over the perturbative contribution coming from high energies (because it is proportional to smaller power of $\lambda$). Likewise, we can obtain the scaling of heat (\ref{q_scaling}) by noting that the low energy excitations carry energy of the order of $\Delta$.  Thus their contribution is $Q\sim n_{\rm ex}\Delta\, \sim |\lambda|^{(d+z)\nu}$. When the exponent $(d+z)\nu$ becomes more than two this low energy contribution becomes sub-leading and the perturbative quadratic scaling is restored to leading order in $\lambda$ (as we mentioned singular non-analytic terms can still persist to higher order in $\lambda$). There is a related and very intuitive way of deriving the scaling (\ref{n_scaling}) adopting the arguments of Kibble and Zurek~\cite{kz1, kz2} to the situation of sudden quenches. Namely, we can interpret the scale $\xi$ as a typical distance between generated quasi-particles. Then the quasi-particle density is
$1/\xi^d\sim |\lambda|^{d\nu}$.

It is interesting that Eq.~(\ref{p_ex}) can also be understood purely by symmetry arguments. Let us assume that the phase corresponding to finite $\lambda$ is characterized by some broken symmetry. The symmetry becomes well defined when $\xi(\lambda)\sim 1/|\lambda|^\nu$ becomes less or comparable to the system size: $\xi(\lambda)\lesssim L$. This defines the minimal quench amplitude $\lambda^\ast\sim 1/L^{1/\nu}$ at which the symmetry is formed. Since the critical point does not correspond to any broken symmetry we anticipate that the overlap between ground state wave-functions $|\psi(0)\rangle$ and $|\psi(\lambda)\rangle$ vanishes for $\lambda\gtrsim \lambda^\ast$. This implies that
\be
\chi_f(0)L^d(\lambda^\ast)^2\gtrsim 1
\label{sym}
\ee
 or equivalently $\chi_f(0)\gtrsim L^{2/\nu-d}$, which is actually the correct scaling. In the case, $d\nu>2$, Eq.~(\ref{sym}) is satisfied even if $\chi_F(\lambda)$ approaches a constant independent of the system size at the critical point. Note that the argument, in principle, allows $\chi_f(0)$ to vanish at the QCP for $d\nu>2$ in the thermodynamic limit. However, as we discussed above, generally one can anticipate that $\chi_f(0)$ is nonzero due to high-energy non-universal contributions not sensitive to the presence of the critical point.

Let us now consider a somewhat different setup, where we still start at the critical point but instead of suddenly quenching the parameter $\lambda$ we assume that it gradually increases over time as a power law
$\lambda(t)=\upsilon t^r/r!\Theta(t)$, where the factor of $1/r!$ is added for convenience. Here $\Theta(t)$ is the step function and $\upsilon$ is a real constant, which controls the proximity to the adiabatic limit. For linear quenches $r=1$ the parameter $\upsilon$ plays the velocity of the quench, for quadratic quenches $r=2$ it is the acceleration and so on. It is interesting that in the limit $r\to 0$ this parameter $\upsilon$ becomes the quench amplitude. For simplicity we also assume that for $r>0$ the final value of $\lambda$ is sufficiently far from the critical point. As in the case of sudden quenches, it is instructive first to perform perturbative in $\upsilon$ analysis of the new state. Using adiabatic perturbation theory~\cite{claudia_kolk, ortiz_2008}, which we will briefly discuss in the next section, one can show that to leading order in $\upsilon$ the probability to occupy the state $|n\rangle$ in the instantaneous (co-moving basis) is
\be
|\delta\psi_n|^2\approx \upsilon^2 {\left|\langle n|V|0\rangle\right|^2\over (E_n-E_0)^{2r+2}},
\label{pn_delta}
\ee
where as before all matrix elements and energies are evaluated at the critical point. We note that the same expression applies to the opposite situation, where one starts far from the critical point and changes the coupling linearly in time until the critical point is reached. This expression can be also generalized to situations when both initial and final couplings are finite~\cite{claudia_kolk}.  Then the total probability to excite the system reads
\be
P_{\rm ex}(\upsilon)\approx \upsilon^2\sum_n {\left|\langle n|V|0\rangle\right|^2\over (E_n-E_0)^{2r+2}}=\upsilon^2 L^d \chi_{2r+2}(0),
\label{g_t}
\ee
where $\chi_{2r+2}(\lambda)$ is the generalized adiabatic susceptibility introduced in the previous section~(\ref{chim}). One can similarly introduce susceptibilities describing the heat density $Q$ and the quasiparticle density $n_{\rm ex}$ for slow quenches. For example, it is easy to see that if the quench ends in the critical point then
\be
Q(\upsilon) \approx \upsilon^2 \chi_{2r+1}(0).
\label{Q_t}
\ee
If, on the other hand, the quench ends far from the critical point then the energy of the system and the number of quasi-particles  are evaluated in the states adiabatically connected to the states $|n\rangle$ at the critical points, thus the expressions for $Q$ and $n_{\rm ex}$ become in general nonuniversal. The situation simplifies in integrable models, where e.g. the quasi-particle number is conserved during the adiabatic evolution. The same is true in nonintegrable models if the excitations are topological and have very slow relaxation. The ambiguities of choosing the basis also disappear if we consider a cyclic process where coupling first linearly increases in time, then saturates, and then decreases back towards its original value, e.g. $\lambda(t)\propto t^r(t_f-t)^r\Theta(t(t_f-t))$.  In this case, the corresponding susceptibilities essentially will have an extra factor of two. Also, there are no complications with choosing the basis if we consider a reverse process where we stop at the critical point.

Now we are in the position of more closely analyzing the scaling of $P_{\rm ex}$ and other quantities. Like in the case of instantaneous quenches $\chi_{2r+2}(\lambda)$ for $r>0$ can diverge at the critical point because of low energy contributions to Eq.~(\ref{g_t}). Since as we discussed earlier ${\rm dim}\chi_{2r+2}=d-2/\nu-2zr$ we see that this is indeed the case for $d<2/\nu+2zr$. As in the case of sudden quenches this divergence leads to super-extensive scaling for the excitation probability:
\be
P_{\rm ex}(\upsilon)\sim |\upsilon|^2 L^{2/\nu+2zr}.
\label{p_ex_r}
\ee
This probability becomes close to one at the crossover rate $\upsilon^\ast\sim 1/L^{1/\nu+zr}$. For faster quenches, $\upsilon>\upsilon^\ast$, the excitation probability is no longer informative and we need to consider other quantities. If we assume that the quasi-particles are well defined then in the case $|\upsilon|>|\upsilon^\ast|$ we can extend Eq.~(\ref{g_t}) as we did for sudden quenches, namely evaluating $\chi_{2r+2}$ at some characteristic coupling slightly away from the critical point. Noting also that $\nu^{-1}={\rm dim}[\lambda]={\rm dim}[\upsilon]-zr$ we see that the scaling dimension of $\upsilon$ is ${\rm dim}[\upsilon]=zr+1/\nu$. This means that the characteristic value of $\lambda$, which should enter the susceptibility is related to $\upsilon$ via $\lambda\sim \upsilon^(1+z\nu r)$. Therefore we recover that for $|\upsilon|>|\upsilon^\ast|$
\be
n_{\rm ex}\sim |\upsilon|^{d\nu\over {z\nu r+1}},
\label{nex1}
\ee
which for linear quenches $r=1$ is indeed the correct scaling first suggested in Refs.~\cite{ap_adiabatic, zurek_adiabatic} and later generalized to nonlinear quenches in Refs.~\cite{sengupta2008, barankov2008}. If we assume that in the final state the spectrum is gapless and characterized by the exponent $z$ (e.g. if the final state of the evolution corresponds to the critical point), we find the the heat is also universal. In particular for $|\upsilon|>|\upsilon^\ast|$:
\be
Q\sim |\upsilon|^{(d+z)\nu\over z\nu r+1}.
\label{qex}
\ee
Equation~(\ref{qex}) is also the correct scaling first suggested in Ref.~\cite{pg_np} for $r=1$. As in the case of sudden quenches these scaling results are expected to be valid only if the corresponding exponents in the powers of $\upsilon$ are smaller than two since otherwise the corresponding susceptibilities $\chi_{2r+2}(\lambda)$ or $\chi_{2r+1}(\lambda)$ are not divergent generically having only cusp singularities at the QCP. In this case the corresponding non-analytic asymptotics of $P_{\rm ex}$, $Q$, and $n_{\rm ex}$ become subleading.

There is also a very intuitive explanation of the scaling (\ref{nex1}). As we mentioned in the previous section, critical points are characterized by the quasi-particle energy scale $\Delta\sim |\lambda|^{z\nu}=|\upsilon t|^{z\nu}$ (we note that the scale $\Delta$, relevant to our discussion, is always associated with quasi-particle excitations, even if quasi-particles are ill defined; many-body energy levels are generally exponentially close to each other). If this energy scale changes sufficiently slowly in time then the energy levels have time to adjust to this change and
adiabatically evolve. However, if $\Delta$ changes sufficiently fast then the adiabaticity breaks down and the states are excited. To find the crossover energy scale separating adiabatic and diabatic states we can use the simple Landau-Zener-Majorana-St\"{u}ckelberg (LZMS) criterion $d\tilde\Delta/dt\sim \tilde\Delta^2$. Using two facts that $\Delta\propto |\lambda|^{z\nu}$ and $\lambda\sim \upsilon t^r$ we find that $\tilde \Delta\sim
|\upsilon|^{z\nu/(z\nu r+1)}$. This scale corresponds to the characteristic momentum $\tilde k\sim |\upsilon|^{\nu/(z\nu r+1)}$ and the characteristic length scale $\tilde\xi\sim 1/\tilde k$. The number of excited quasi-particle states in the spatially uniform system is then $n_{\rm ex}\sim \tilde k^d\sim |\upsilon|^{d\nu/(z\nu r+1)}$, which is exactly as in Eq.~(\ref{nex1}). We note that in this form the argument does not require that the initial or final point of the evolution coincides with the critical point. It is only important that the QCP is crossed during the time
evolution. In the next section, in which we discuss adiabatic perturbation theory, we will show that this is indeed the case. The crossover to the quadratic scaling is also expected to be generic. In general in the expressions (\ref{g_t}) and (\ref{Q_t}) one has to evaluate the corresponding susceptibilities in the initial and the final points of the evolution. Away from the phase transitions we expect no singularities in the corresponding susceptibilities and thus the quadratic scaling will hold. This simple scaling argument can be also reformulated in the
spirit of Kibble-Zurek mechanism~\cite{kz1, kz2}. Namely, one can interpret the length scale $\tilde\xi$ as a characteristic distance between generated quasi-particles and the inverse energy scale $\tilde \Delta^{-1}$ as a time scale where the process is not adiabatic and the defects are generated.

{\em Example: transverse-field Ising model and multi-dimensional extensions.} Let us now illustrate how the scalings we derived apply to the specific example we introduced in Sec.~\ref{sec:review}. We start from sudden quenches. We also focus on the thermodynamic limit $L\to\infty$ and $\lambda$ is independent of $L$. Using
the explicit structure of the ground state wave function~(\ref{gs_tfim}) it is easy to check that the overlap of two
different ground states corresponding to a particular pair of fermions with momenta $k,-k$ is
\be
\langle \Psi_0^k(\lambda)|\Psi_0^k(\lambda')\rangle=\cos\left[(\theta_k-\theta_k^\prime)/ 2\right],
\ee
where $\theta_k$ and $\theta_k^\prime$ correspond to the couplings $\lambda$ and $\lambda'$ respectively. This implies that the probability of exciting a pair of quasi-particles with momenta $k$ and $-k$ by quenching the parameter $\lambda'$ to $\lambda$ is $p_{\rm ex}(k)=\sin^2\left([\theta_k-\theta_k^\prime]/2 \right)$.
Noting that we are interested in the limit $\lambda'=0$ and using low energy expressions for $\theta_k$ and
$\theta_k^\prime$ we find
\be
p_{\rm ex}(k)\approx {1\over 2}\left[1-{|k|\over \sqrt{k^2+\lambda^2}}\right].
\ee
The density of quasi-particles excited in the quench is then obviously obtained by integrating $p_{\rm ex}(k)$ over different momenta
\be
n_{\rm ex}\approx \int_{-\pi}^{\pi} {dk\over 2\pi} p_{\rm ex} (k)\approx {|\lambda|\over 2\pi}\,.
\ee
This scaling indeed agrees with our general expectation $n_{\rm ex}\sim |\lambda|^{d\nu}$, noting that $d=\nu=1$. Similarly we can find heat, noting that each quasi-particle with the momentum $k$ carries energy $\epsilon_k\approx 2\sqrt{\lambda^2+k^2}$. Then
\be
Q\approx 2\int_{-\pi}^\pi {dk\over 2\pi}\sqrt{k^2+\lambda^2}p_{\rm ex}(k) \approx {1\over 2\pi}\lambda^2 \ln{2\pi\over|\lambda|},
\ee
which also agrees with $Q\sim |\lambda|^{(d+z)\nu}$. However, because in this case the power $(d+z)\nu$ is exactly equal to two, we find an additional logarithmic dependence on both $\lambda$ and the cutoff $\pi$. The logarithmic dependence is natural at the point where we expect crossover from the exponent two at $(d+z)\nu>2$ to the exponent less than two in the opposite case. Even though the transverse-field Ising model is defined only in 1D, one can extend it to higher dimensions by formally considering the free-fermion Hamiltonian (\ref{h_ising1}) in higher dimensional lattices. E.g. in 2D a very similar Hamiltonian represents the fermionic sector of the Kitaev model~\cite{kitaev} or describes electrons and holes in graphene~\cite{antonio_rmp}. Then it is easy to check that $n_{\rm ex}$ has quadratic scaling with $\lambda$ in 2D and above with extra logarithmic corrections in 2D where $d\nu=2$. The heat has quadratic scaling above 1D (see Ref.~\cite{quench_qcp} for additional details).

The transverse-field Ising model can be also solved for slow quenches. For simplicity we will analyze only the linear dependence $\lambda=\upsilon t$. Note that the wave function factorizes into a direct product of states corresponding to different momenta with either zero fermions or two fermions in each state. Thus the dynamical problem factorizes into a direct sum of LZMS problems.   If the magnetic field linearly crosses the QCP then the transition probability is given by~\cite{dziarmaga}
\be
p_{\rm ex}(k)\approx \exp(-2\pi k^2/\upsilon)
\ee
The density of quasi-particles generated in such process is obtained by integrating $p_{\rm ex}(k)$ over different momentum states yielding in the slow limit $n_{\rm ex}\approx \sqrt{\upsilon}/(2\pi\sqrt{2})$. This is indeed the expected scaling $n_{\rm ex}\sim |\upsilon|^{d\nu/(z\nu+1)}$ for a particular set of exponents $d=\nu=z=1$. The problem can be also solved for the linear quench starting at the critical point giving identical scaling with
a slightly different prefactor~\cite{claudia_kolk}. To see the crossover to the quadratic scaling one needs to consider extension of the Hamiltonian~(\ref{h_ising1}) to higher dimensions. In this case for each momentum state we are effectively dealing with half LZMS problems, i.e. LZMS problems where the initial
coupling corresponds to the minimum gap. Asymptotically, in the slow limit $\upsilon\ll \epsilon_k^2$, this probability is found from Eq.~(\ref{pn_delta}):
\be
p_{\rm ex}(k)\approx (1024)^{-1}\upsilon^2/k^4
\ee
It is clear that above four dimensions the density of quasi-particles $n_{\rm ex}\sim \int d^dk p_{\rm ex}$ will be
quadratic, dominated by excitations to high energies of the order of the cutoff. Below four dimensions the integral converges at large $k$ so transition probabilities are dominated by small momenta $k\sim\sqrt{\upsilon}$ and the scaling $n_{\rm ex}\sim |\upsilon|^{d\nu/(z\nu+1)}=|\upsilon|^{d/2}$ is restored.

\section{Adiabatic Perturbation Theory}

\subsection{Sketch of the Derivation}

In this section we will present a more accurate derivation of the scaling of various quantities near the critical point. In particular, our aim is to derive the expression for the transition probability in the leading order in $\upsilon$. A very convenient framework to analyze these scaling laws is given by adiabatic perturbation theory. As we will see this approach will allow us not only to reproduce Eq.~(\ref{pn_delta}), but also extend this result to the regime where the corresponding susceptibility $\chi_{2r+2}(\lambda)$ diverges at the QCP and justify Eqs.~(\ref{nex1}) and (\ref{qex}). In Ref.~\cite{ap_adiabatic} this theory was originally applied to derive the scaling of the density of quasi-particles~(\ref{nex1}). In Ref.~\cite{pg_np} it was shown that this approach correctly predicts the crossover between analytic and non-analytic regimes of scaling of heat (excess energy) with the rate
$\upsilon$ for linear quenches. In Refs.~\cite{claudia_kolk, quench_qcp} this theory was extended to sudden quenches near critical points and also reproduced correct scaling results. While adiabatic perturbation theory is not quantitatively accurate in non-analytic regimes (where the response is not quadratic)~\cite{ap_adiabatic, pg_np}, i.e., it does not correctly reproduce the prefactor, it predicts correct scaling laws in many different situations. The only known exceptions are in the so-called non-adiabatic regime, where the system size or other macroscopic length scale enters the scaling of heat or quasi-particle density~\cite{pg_np, quench_qcp}. Such regimes can appear e.g. if we are dealing with low-dimensional systems, which have low energy bosonic excitations, especially at finite temperatures, where the violation of scaling (\ref{nex1}) comes from the overpopulation of low-energy modes. This regime is quite special and we will not consider it here (we refer the reader interested in more details to Refs.~\cite{pg_np, quench_qcp}).

In the beginning of the section we will closely follow the discussion of Refs.~\cite{claudia_kolk, ortiz2008}.  We consider a very general setup where the system is described by a Hamiltonian $H(t)= H_0+\lambda(t)V$, where $\lambda(t)$ monotonically changes in time between initial and final values $\lambda_i$ and $\lambda_f$. For simplicity in this chapter we will focus only in linear dependence $\lambda(t)=\upsilon t$, though this assumption is not important. The limit $\upsilon\to\infty$ corresponds to a sudden quench and $\upsilon\to 0$ to a slow quench. We will always assume that for sudden quenches $\lambda_f$ is close to $\lambda_i$, while for slow quenches this condition is not necessary. In both cases adiabatic perturbation theory will be justified by the proximity of the system to the ground state after the quench.

Our goal is to approximately solve the  Schr\"odinger equation
\be
i\partial_t |\psi\rangle =H(t) |\psi\rangle,
\label{schr_eq}
\ee
where $|\psi\rangle$ is the wave function. It is convenient to
rewrite Eq.~(\ref{schr_eq}) in the adiabatic (instantaneous) basis:
\be
|\psi(t)\rangle=\textstyle\sum_n a_n(t) |n(t)\rangle,\quad  H(t) |n(t)\rangle=E_n(t)|n(t)\rangle,
\ee
where $E_n(t)$ are the instantaneous eigenvalues. The eigenstates $|n(t)\rangle$ implicitly depend on time through the coupling $\lambda(t)$. Substituting this expansion into the Schr\"odinger equation and multiplying it by $\langle m|$ (to simplify our notations we drop the time label $t$ in $|n(t)\rangle$) we find:
\be
i\partial_t a_n(t)+i\textstyle\sum_m a_m(t) \langle n|\partial_t|m\rangle=E_n(t) a_n(t).
\ee
Next we will perform a unitary transformation:
\be
a_n(t)=\alpha_n(t) \exp\left[-i\Theta_n(t)\right],\quad
\Theta_n(t)=\textstyle\int_{t_i}^t E_n(\tau)d\tau.
\ee
The lower limit of integration in the expression for $\Theta_n(t)$ is arbitrary. We chose it to be equal to $t_i$ for convenience. Then the Schr\"odinger equation becomes
\be
\dot\alpha_n(t)=-\textstyle\sum_m \alpha_m(t) \langle n|\partial_t|m\rangle\exp\left[i(\Theta_n(t)-\Theta_m(t))\right].
\ee
In turn this equation can be rewritten as an integral equation
\be
\alpha_n(t)=-\textstyle\int_{t_i}^t dt' \textstyle\sum_m \alpha_m(t') \langle n|\partial_{t'}|m\rangle\mathrm e^{i(\Theta_n(t')-\Theta_m(t'))}.
\label{int_eq}
\ee
If the energy levels $E_n(t)$ and $E_m(t)$ are not degenerate, the matrix element $\langle n|\partial_t|m\rangle$ can be written as
 \be
 \langle n|\partial_t|m\rangle=-{\langle n| \partial_t  H|
 m\rangle\over E_n(t)-E_m(t) }=-\dot\lambda(t){\langle n| V| m\rangle\over
 E_n(t)-E_m(t) },
 \ee
If $\lambda(t)$ is a monotonic function of time, like in our case, then in Eq.~(\ref{int_eq})
one can change variables from $t$ to $\lambda(t)$ and derive
\be
\alpha_n(\lambda)=-\textstyle\int_{\lambda_i}^\lambda d\lambda' \textstyle\sum_m \alpha_m(\lambda')
\langle n|\partial_{\lambda'}|m\rangle\mathrm e^{i(\Theta_n(\lambda')-\Theta_m(\lambda'))},
\label{int_eq1}
\ee
where
\be
\Theta_n(\lambda)=\textstyle\int_{\lambda_i}^\lambda d\lambda' E_n(\lambda')( \dot\lambda')^{-1}.
\ee
Equations~(\ref{int_eq}) and (\ref{int_eq1}) suggest a systematic
expansion in the transition amplitudes to the excited states. We are interested in either the limit $\dot\lambda=\upsilon\to 0$, which suppresses transitions because of the highly oscillating phase factor, or
the limit of small $|\lambda_f-\lambda_i|$, where the transitions are suppressed by the smallness of the integration domain. To leading order in adiabatic perturbation theory only the diagonal terms with $m=n$ should be retained in the sums in Eqs.~(\ref{int_eq}) and (\ref{int_eq1}). These terms result in the emergence of a Berry phase~\cite{shankar}:
\be
\Phi_n(t)=-i\textstyle\int_{t_i}^t dt' \langle n
|\partial_{t'}|n\rangle=-i\textstyle\int_{\lambda_i}^{\lambda(t)} d\lambda' \langle n
|\partial_{\lambda'}|n\rangle\,,
\ee
so that $a_n(t)\approx a_n(0)\exp[-i\Phi_n(t)]$. In general, the Berry phase can be incorporated into our formalism by performing a unitary transformation $\alpha_n(t)\to \alpha_n(t)\exp[-i\Phi_n(t)]$ and changing $\Theta_n\to \Theta_n+\Phi_n$ in Eqs.~(\ref{int_eq}) and (\ref{int_eq1}).

In many situations, when we deal with real Hamiltonians, the Berry phase is identically equal to zero. However, in some cases when more than one coupling constant change in time, the contribution of the geometric phases can be important, so that it can change the results for scaling of the physical quantities near the phase transition~\cite{tomka}. Here we only note that since the geometric phase is related to the topology of the phase space, the evolution of physical quantities can depend on the path in the parameter space. The effects of the geometric phase are enhanced near the diabolic points corresponding to the level crossings. Geometric phase effects can be important for open systems, which effectively can be modeled by non-Hermitean (complex) Hamiltonians.

Assuming that the geometric phase is not important, let us compute the first order correction to the wave function assuming that initially the system is in the pure state $n=0$, so that $\alpha_0(0)=1$ and $\alpha_n(0)=0$ for $n\neq 0$. To leading order in $\dot\lambda$ we can keep only one term with $m=0$ in the sums in Eqs.~(\ref{int_eq}) and~(\ref{int_eq1}) and derive
\be
\alpha_n(t)\approx-\textstyle\int_{t_i}^t dt' \langle n|\partial_{t'}|0\rangle\mathrm e^{i[\Theta_n(t')-\Theta_0(t')]}.
\label{int_eq2}
\ee
 or alternatively
 \be
 \alpha_n(\lambda)\approx-\textstyle\int_{\lambda_i}^\lambda d\lambda' \langle
 n|\partial_{\lambda'}|0\rangle\mathrm
 e^{i[\Theta_n(\lambda')-\Theta_0(\lambda')]}.
 \label{int_eq3}
 \ee
The transition probability from the level $|0\rangle$ to the level
$|n\rangle$ as a result of the process is determined by
$|\alpha_n(\lambda_f)|^2$. Let us note that in the limit of $\dot\lambda=\upsilon \to 0$ one can expect two types of contributions to Eq.~(\ref{int_eq3}). (i) A non-analytic contribution comes from the saddle
points of the phase difference $\Theta_n(\lambda)-\Theta_m(\lambda)$, which in turn correspond to
the complex roots of $E_n(\lambda)=E_0(\lambda)$. These terms result in
exponential dependence of the transition probability on rate
$|\alpha|^2\sim \exp[-A/|\dot\lambda|]$ like in the usual LZMS
problem (see Ref.~\cite{claudia_kolk} for the additional discussion).
(ii) An analytic contribution comes from the moments where we turn on and turn off the process. This second contribution, to leading order in $\dot\lambda$, can be obtained by integrating
Eq.~(\ref{int_eq3}) by parts~\cite{claudia_kolk, ortiz_2008}:
\be
\alpha_n(\lambda_f)\approx \left[i \dot\lambda{\langle n|
\partial_\lambda|0\rangle\over E_n(\lambda)-E_0(\lambda)}\right]
\mathrm e^{i(\Theta_n(\lambda)-\Theta_0(\lambda))}\Biggr|_{\lambda_i}^{\lambda_f}\,.
 \label{expansion1}
 \ee
From this we find the analytic part of the transition probability
\beq
|\alpha_n(\lambda_f)|^2&\approx& \upsilon^2\left[
{|\langle n|\partial_{\lambda_i}|0\rangle|^2\over (E_n(\lambda_i)-E_0(\lambda_i))^2}
+{|\langle n|\partial_{\lambda_f}|0\rangle|^2\over (E_n(\lambda_f)-E_0(\lambda_f))^2}\right]\nonumber\\
&-&2\upsilon^2 {\langle n|\partial_{\lambda_i}| 0\rangle\over E_n(\lambda_i)-E_0(\lambda_i)}
{\langle n|\partial_{\lambda_f}| 0\rangle\over E_n(\lambda_f)-E_0(\lambda_f)}\cos\left[\Delta\Theta_{n0}\right],
\label{tr_prob}
\eeq
where $\Delta\Theta_{n0}=\Theta_n(\lambda_f)-\Theta_0(\lambda_f)-\Theta_n(\lambda_i)+\Theta_0(\lambda_i)$.
Usually if we deal with many levels the last fast-oscillating term will average out to zero. The remaining first two terms are the transition probabilities associated with turning on and off the coupling $\lambda$. If only the first term dominates the scaling (e.g. because the initial state corresponds to a QCP) we recover Eq.~(\ref{pn_delta}) for $r=1$. If we deal with nonlinear quenches such that $\dot\lambda=0$ at the initial and final points of the evolution then we need to continue integrating Eq.~(\ref{int_eq3}) by parts until we hit the first non-vanishing time-derivative of $\lambda(t)$. It is straightforward to see that in this way we reproduce Eq.~(\ref{pn_delta}) for any integer power $r$. The resulting expression can be analytically continued to all positive $r$. Interestingly it correctly reproduces the result of ordinary perturbation theory for sudden quenches $r=0$ where the parameter $\upsilon$ plays the role of the quench amplitude.

\subsection{Applications to Dynamics near Critical Points}

Let us now perform the scaling analysis of Eq.~(\ref{int_eq3}). First we consider the situation of sudden quenches with the initial coupling corresponding to the critical point $\lambda_i=0$ (or alternatively $\lambda_f=0$). We will assume that excitations are dominated by quasi-particles created in pairs with opposite momenta $k,-k$. As we saw in the previous section this assumption is well justified for non-interacting models like the transverse-field Ising
model. In general, for interacting non-integrable models eigenstates consist of quasi-particles dressed by interactions and the quasi-particles acquire finite life time. If the quasi-particle nature does not qualitatively change because of the interactions we expect that the scaling of the quasi-particle density will remain the same as the scaling of the excitation probability.In fact, the whole scaling analysis can be performed in the spirit of Sec.~\ref{sec:scaling}, which does not rely on the assumption of having well-defined quasi-particles. However, such an assumption makes all derivations more transparent.

Assuming that $|n\rangle$ corresponds to the quasi-particle pair with momenta $k,-k$ (to simplify our notation we denote such pair states by $|k\rangle$ and the corresponding energy of the pair by $\epsilon_k$) we find
\be
\alpha_k(\lambda_f)\approx -\int_{0}^{\lambda_f} d\lambda {\langle k|V|0\rangle\over \epsilon_k(\lambda)-\epsilon_0(\lambda)}.
\ee
Since the scaling dimension of the operator $\lambda V$ is the same as the scaling dimension of energy we expect the following scaling of the matrix element:
\be
{\langle k|V|0\rangle\over \epsilon_k(\lambda)-\epsilon_0(\lambda)}={1\over \lambda} G(k/|\lambda|^\nu),
\ee
where $G(x)$ is some scaling function. We anticipate that in the limit $x\gg 1$ we have $G(x)\sim x^{-1/\nu}$ so that the matrix element becomes independent of $\lambda$ at $k\gg \lambda^\nu$. In the opposite limit, $x\ll 1$, the scaling function can either saturate if there is a gap or vanish as some power of $x$ if there is no gap.
It is easy to check that for the transverse-field Ising model the scaling assumption is indeed satisfied with $G(x)\propto x/(x^2+1)$. This scaling ansatz immediately allows us to analyze the behavior of the quasi-particle density:
\be
n_{\rm ex}\approx {1\over L^d}\textstyle\sum_k |\alpha_k|^2.
\label{nex2}
\ee
Then taking the thermodynamic limit where the sum over $k$ becomes an integral and changing variables
$\lambda\to\lambda_f\eta$ and $k\to |\lambda_f|^{\nu}\xi$ we
immediately find
\be
n_{\rm ex}(\lambda_f)\approx |\lambda_f|^{d\nu}\int {d^d\xi\over (2\pi)^d} \left|\int_0^{1} d\eta{1\over \eta} G(\xi/|\eta|^{\nu})\right|^2.
\ee
This expression gives the right scaling (\ref{n_scaling}) provided that the integral over $\xi$ converges at large $\xi$. The convergence clearly depends on the large $\xi$ asymptotics of the function $G(\xi/|\eta|^{\nu})\sim 1/|\xi|^{1/\nu}$. The integral over $\xi$ clearly converges provided that $d\leq 2/\nu$ or $d\nu\leq 2$. In
the opposite case the integral over momenta is dominated by large $k\gg |\lambda_f|^{\nu}$. In this case Eq.~(\ref{nex2}) reduces to the perturbative quadratic result, e.g. $n_{\rm ex}(\lambda_f)\approx
2\lambda_f^2\chi_f(0)$, where the factor of two reflects that each excited state contributes to two quasi-particles. Note that the assumption of scale independence of the transition matrix element is equivalent to the
assumption that $\chi_f$ does not vanish at the critical point for $d\nu>2$. As we pointed out earlier, this is generally expected because transitions to the high energy states are insensitive to the proximity
to the critical point dominate excitations in the system (see detailed discussion in Ref.~\cite{quench_qcp} for a particular sine-Gordon model). Likewise, one can derive the correct scaling for the heat density (\ref{q_scaling})
provided that $(d+z)\nu<2$; one reproduces the quadratic perturbative result in the opposite case.

In a similar fashion we can derive the results for the adiabatic case, when the coupling constant changes slowly in time. For simplicity we focus again on linear quenches $\lambda(t)=\upsilon t$. For now it will not be important whether we start (finish) at the critical point or the initial and final couplings are on different sides of the QCP. The expression for the transition amplitude then becomes
\be
\alpha_k(\lambda_f)\approx -\int_{\lambda_i}^{\lambda_f} d\lambda
{\langle k|V|0\rangle\over \epsilon_k(\lambda)-\epsilon_0(\lambda)}
\exp\left[{i\over\upsilon}\int_0^\lambda d\lambda' (\epsilon_k(\lambda')-\epsilon_0(\lambda'))\right].
\ee
Let us now introduce another scaling function $F(x)$ according to $\epsilon_k(\lambda)-\epsilon_0(\lambda)=|\lambda|^{z\nu}
F(k/|\lambda|^\nu)$ with the asymptotic power law $F(x)\sim |x|^z$ for $|x|\gg
1$ (the small $x$ limit depends on whether the system is gapless or not). A convenient change of variables is
$k=|\upsilon|^{\nu/(z\nu+1)}\eta$ and $\lambda=|\upsilon|^{1/(z\nu+1)}\eta$. Then the expression for the density of quasi-particles becomes
\be
n_{\rm ex}\sim \int {d^dk \over (2\pi)^d }|\alpha_k|^2=|\upsilon|^{d\nu\over
z\nu+1} \int {d^d\eta\over (2\pi)^d} |\alpha(\eta)|^2,
\label{nex3}
\ee
where
\be
\alpha(\eta)=\int_{\xi_i}^{\xi_f} d\xi {1\over
\xi}G\left({\eta\over\xi^{\nu}}\right)\exp\left[i\int_{\xi_i}^\xi d\xi_1
\xi_1^{z\nu}F(\eta/\xi_1^{\nu})\right].
\label{alpha}
\ee
An analogous expression is found for heat.  If $\upsilon$ is small then the limits of integration over $\xi$ can
be extended to $(-\infty,\infty)$ if $\lambda_i<0$ and $\lambda_f>0$ or to $(0,\infty)$ if $\lambda_i=0$ and $\lambda_f$ is finite. Indeed, the integral over $\xi$ is always convergent because of the highly oscillating exponent. If additionally the integral over the rescaled momentum $\eta$ can be extended to $\infty$ we get the
desired scaling for the quasi-particle density (\ref{nex1}).

To analyze the convergence (assuming for simplicity that $|\lambda_f|\gg |\lambda_i|$ and $\lambda_i$ is close to the QCP) we note that at large $k$ the asymptotic expression for the transition amplitude $\alpha(k)$ according to Eq.~(\ref{expansion1}) is
\be
\alpha(k)\approx i\upsilon {\langle k| V|0\rangle\over (E_k-E_0)^2}\sim {\upsilon\over k^{z\nu+1/\nu}},
\ee
where we kept only the term corresponding to the initial coupling $\lambda_i=0$ because it describes the singularity. We thus immediately infer that the integral over the momentum $k$ (and thus over the rescaled momentum $\eta$) can be extended to $\infty$ (so that the scaling (\ref{nex3}) is valid) as long as $d\leq 2z\nu+2/\nu$ or equivalently $d\nu/(z\nu+1)\leq 2$. In the opposite case, the usual adiabatic perturbation theory is restored and we recover the quadratic perturbative result $n_{\rm ex}\sim \upsilon^2(\chi_f(\lambda_i)+\chi_f(\lambda_f)$. We expect this scaling to be valid even if both $\lambda_i$ and $\lambda_f$ are away from the quantum critical region. Indeed, in this case one does not expect any divergences in susceptibilities $\chi_m(\lambda)$ at both $\lambda_i$ and $\lambda_f$ so the quadratic scaling should dominate the dynamics over the higher critical power $|\upsilon|^{d\nu/(z\nu+1)}$ emerging
from the scaling argument. A somewhat special situation emerges if the relevant susceptibilities $\chi_m(\lambda)$ are very small at $\lambda_i$ and $\lambda_f$. Then one might expect that the scaling $|\upsilon|^{d\nu/(z\nu+1)}$ will be applicable even when the exponent exceeds two. A similar situation occurs if the initial and final couplings are finite but the time dependence of $\lambda(t)$ is smooth near $t_i$ and $t_f$ and linear only near the
critical point. These expectations are well justified only for non-interacting systems like the transverse-field Ising model, where dynamics can be mapped to independent LZMS transitions. In general, interactions lead to dephasing of quasi-particles, which is equivalent to resetting the dynamical process each time the quasi-particle phase is lost. Then one can expect the quadratic scaling with $\upsilon$ will be again restored for $d\nu/(z\nu+1)\geq
2$. At the moment this issue remains an open problem. We note that in the case $d\nu/(z\nu+1)$ one can expect that the non-analytic term coming from low energies will still survive; it will just become sub-leading in the limit of small $\upsilon$.

Another important point is that the susceptibilities $\chi_{2r+2}(\lambda)$ for $r>0$ can diverge even away from QCP as long as the system is gapless. Indeed e.g. for bosonic theories gapless phases correspond to $\nu\to\infty$. Then the susceptibility $\chi_{2r+2}$ diverges for $d<2zr$. This means that even in gapless noncritical phases in low dimensions the quadratic scaling can be violated and replaced by $n_{\rm ex}\sim |\upsilon|^{d/zr}$ and
$Q\sim |\upsilon|^{(d+z)/zr}$. This is consistent with general prediction of less adiabatic response in low dimensions~\cite{pg_np}. It is also interesting to note that in the noncritical case for $d/zr<2$ the density of quasiparticles (or for $(d+z)/zr<2$ the heat density) depend only on the total time of the quench, but not its shape. Indeed since the total quench time $\tau\sim 1/\upsilon^{1/r}$ we see that $n_{\rm ex}\sim |\upsilon|^{d/zr}\sim (1/\tau)^{d/z}$, i.e. independent of $r$. On the other hand for $d/zr>2$ and in gapped phases in arbitrary dimensions $n_{\rm ex}\sim (1/\tau)^{2r}$ is very sensitive to the exponent $r$ and thus to the shape of the quench. The existense of these two regimes is true in other situations as well. Thus in Ref.~\cite{kollar_09} the two regimes of sensitivity of the energy (heating) to the shape pulse were discovered for interacting gapless fermionic systems. In general relevance of the ramping protocol to the non-adiabatic dynamics in interacting many-body systems remains an open problem, which is likely very important for understanding the problem of dissipation.

\subsection{Quenches at Finite Temperatures, and the Role of Quasi-particle Statistics}

So far, we have exclusively focused on the situation where the system is
prepared in the ground state. A natural question arises of how one can extend these results to the finite-temperature domain. It is well established that in equilibrium near a QPT influence of the QCP extends into finite temperatures in the so-called quantum critical region~\cite{sachdev_book} which can be quite extensive. We can anticipate that a similar situation occurs for non-equilibrium systems. Let us point out that the definition of temperature can be somewhat ambiguous away from equilibrium. One can consider two natural setups where the system is coupled to some external reservoir even during a dynamical process (see e.g. Ref.~\cite{silva2009}). In this situation results are sensitive to the coupling strength to the bath and in this sense are not universal. However, one can
imagine a quite different setting, where the system is initially prepared in some thermal state and then the consequent dynamics is purely Hamiltonian, i.e., the effects of the bath are negligible during the time evolution. Such a setup was considered in Refs.~\cite{pg_np, quench_qcp, cardy2009}. In this setup, no coupling to the  environment is assumed during the evolution and one can expect universality of the results. Such setups routinely appear now in the
cold-atom context in which systems, after initial preparation, are essentially isolated from the environment~\cite{bloch_rmp}. They also appear generally in thermodynamics when one considers adiabatic or nearly-adiabatic processes (in which changes happen on time scales faster than the time
of equilibration with the environment)~\cite{LL5}. Universality can only be expected in this setup if the system is
initially prepared close to a QCP. So in this section we restrict our analysis only to such situations.

On general grounds one can expect that at finite temperatures the statistical nature of low energy excitations should play a significant role. Indeed, if we are dealing with bosonic low energy quasi-particles we expect that the initial thermal population enhances the transition probability to the corresponding quasi-particle modes. The reason is that bosonic excitations have a bunching tendency, i.e., a tendency to stimulate transitions to a particular mode if it is
already occupied. On the other hand, if quasi-particles are fermionic then Pauli blocking occurs due to already-occupied modes and fewer quasi-particles can be additionally excited. In principle, one can expect situations where low-energy excitations are either described by quasi-particles with fractional statistics like in Kitaev~\cite{kitaev} and sine-Gordon models~\cite{Lukyanov97}, or by many-particle excitations without well defined statistics. There the effect of finite temperature on dynamics is unknown at present and thus will not be discussed here.

If quasi-particles in the quantum critical region are bosonic then it is straightforward to show that the number of excited quasi-particle pairs in the mode with momentum $q$ at initial temperature $T$ is related to
the number of quasi-particle pairs created at zero temperature via a very simple expression valid for any dependence of
$\lambda(t)$~\cite{pg_np}:
\be
n_{\rm ex}(q,T)=n_{\rm eq}(q,T)+n_{\rm ex}^0(q)\coth(\epsilon_q/2T).
\label{nex_bos}
\ee
In Eq.~(\ref{nex_bos}), $n_{\rm eq}(q,T)$ is the initial equilibrium population of the
bosonic mode described by the Bose-Einstein distribution function,
$\epsilon_q$ is the (initial) energy of the quasi-particle, and $n_{\rm ex}^0(q)$ is the number of quasi-particles created in the same dynamical process if the initial temperature is
zero. At small temperatures $\epsilon_q\gg T$ this expression
clearly reduces to the zero temperature limit. In the opposite high
temperature limit the transition probability is enhanced by
the factor $2T/\epsilon_q\gg 1$. Similarly, in the fermionic case we have~\cite{quench_qcp}
\be
n_{\rm ex}(q,T)=n_{\rm eq}(q,T)+n_{\rm ex}^0(q)\tanh(\epsilon_q/2T).
\label{nex_ferm}
\ee
This expression also reproduces the zero-temperature result for $\epsilon_q\gg T$ while yielding suppression of the transition probability by a factor $\epsilon_q/(2T)$ in the opposite limit. Since the total number of dynamically excited quasi-particles is found by summing $n_{\rm ex}(q,T)-n_{\rm eq}(q,T)$ we see that to find the finite temperature scaling we need to change $d\to d-z$ for bosons and $d\to d+z$ for fermions in universal
expressions~(\ref{n_scaling}) and (\ref{nex1}). The same applies to the expression for heat~(\ref{q_scaling}). For
example, for a slow quench starting at the critical point we find, instead of Eq.~(\ref{nex1}),
\be
n_{\rm ex}^{\rm bos}(T)\sim |\upsilon|^{(d-z)\nu/(z\nu r+1)},\quad n_{\rm ex}^{\rm ferm}(T)\sim |\upsilon|^{(d+z)\nu/(z\nu r+1)}.
\label{nex_t}
\ee
As before, these scalings are generally valid (to leading order in $\upsilon$) only if the
corresponding exponents are less than two; otherwise the scaling of $n_{\rm ex}$, $Q$ with $\upsilon$ becomes quadratic.

There is actually an additional potential issue where the scaling (\ref{nex_t}) can break. This can happen because the integrals over different momentum states $q$ can become divergent at small $q$. This corresponds to emergence in a new non-adiabatic regime of response of the system where the corresponding intensive observables become system-size-dependent~\cite{pg_np}. This situation usually happens in low-dimensional bosonic systems where such infrared divergences correspond to overpopulation of the low energy modes. While there are specific examples of such behavior for slow dynamics in weakly interacting Bose gases~\cite{pg_np} and for dynamics in the sine-Gordon
model near the massive bosonic limit~\cite{quench_qcp} there is no general understanding of this effect.

\section{Going Beyond Condensed Matter}

Here we focus on some possible extensions of the field of application of slow dynamics to other areas of physics, in particular, to cosmology and field (string) theory. The purpose of this section is to show that there are close similarities appearing between quantum dynamics near various space-time singularities and quantum critical dynamics. We do not intend to give a comprehensive overview of new problems nor present details of derivations. We are not experts in these fields. Our purpose is only to show that the applicability of our previous discussion extends well beyond condensed matter and atomic physics. Our first example will be from cosmology: we consider an early stage of expansion of the universe in $d$-dimensional space-time and assume that it is described either by a standard de Sitter-type metric or by its simplest generalization, which includes a slow-roll parameter. The second example arises mainly in some recent studies in string theory, where the question is a propagation of strings on some time-dependent background geometries. In particular, these geometries can be singular. This type of singularity is somewhat reminiscent of a phase transition, but it is different, and is of a completely geometrical nature. However, it can have an interpretation in terms of dynamical systems.  We consider the simplest illustrative example --  quantum evolution in a so-called Milnes universe. In both examples the analogue of a QPT is provided by a singularity in the space-time background on which the quantum system is evolved.

\subsection{Adiabaticity in Cosmology}

A beginning of the Universe can also be considered as a (quantum) phase transition. In modern cosmology one usually considers a scalar field (a Higgs field) on some curved background evolving according to the Einstein equation. A regime of parameters when the potential energy of that field dominates over the kinetic term leads to the
rapid blowing up of the Universe, called inflation. Inflation can be modeled by an exponential scale factor in the metric and is called the de Sitter epoch. The velocity of inflation is given by a Hubble parameter which is assumed to be a constant in the de Sitter model. The natural adiabaticity parameter is then given by the ratio of the
Hubble constant and the mass of the scalar field. The initial point of the expansion corresponds to the singularity in the metric where the scale factor $a(t)$ (see below) is very small. It is therefore natural to look into the cosmological evolution problem from the point of view of critical dynamics near the phase transition.

We consider a spherically-symmetric metric in $(d+1)$-dimensional space-time described by a single time-dependent scale factor
\beq
ds^{2}=-dt^{2}+a(t)^{2}d{\bf x}^{2},
\eeq
where ${\bf x}$ denotes a $d$-vector. The Hubble parameter is $H=\dot{a}(t)/a(t)$, where the ``dot'' denotes a time derivative. We consider a massive scalar field $\Phi({\bf x},t)$ in this geometry, which is minimally coupled to the gravity. The corresponding Klein-Gordon equation is given by~\cite{Linde}
\beq
\ddot\Phi(x,t)+dH\dot{\Phi}(x,t)-\frac{1}{a^{2}(t)}\nabla^{2}\Phi(x,t)+m^{2}\Phi(x,t)=0\,.
\eeq
Introducing the rescaled field $\phi(x,t)=a^{d/2}\Phi(x,t)$ and making a Fourier transform we obtain
\beq
\ddot\phi_{k}(t)+\Omega_{k}^{2}\phi_{k}(t)=0,
\eeq
where $\phi_{k}(t)$ is a Fourier component of $\phi(x,t)$,
\beq
\Omega^{2}_{k}(t)=\omega_{k}^{2}(t)-\left[\frac{d}{2}\left(\frac{d}{2}-1\right)
\left(\frac{\dot{a}}{a}\right)^{2}+\frac{d}{2}\frac{\ddot{a}}{a}\right]=
\omega_{k}^{2}-H^{2}\left[\left(\frac{d}{2}\right)^{2}+\frac{d}{2}\epsilon\right],
\eeq
$\omega_{k}^{2}=m^{2}+k^{2}/a^{2}$, and we introduced a slow-roll parameter $\epsilon=\dot{H}/H^{2}$. This parameter is
usually considered to be small during the inflation stage, thus indicating the smallness of the kinetic energy of a scalar field.

To proceed with quantization, we expand the Fourier components into a time-dependent creation and annihilation operator basis, $\phi_{k}\rightarrow \hat{\phi}_{k}(t)=\psi_{k}(t)\hat a_{k}+\psi^{*}_{k}(t)\hat a^{\dag}_{k}$. The quantization is consistent with the following Klein-Gordon scalar product: {$(\phi_{1},\phi_{2})=i\int
d^{d}x (2\pi)^{-d}[\phi_1^{*}(x,t)\dot{\phi}_2(x,t)-\dot{\phi}_{1}^\star(x,t)\phi_{2}(x,t)]$.
Because of the time-dependence we have to distinguish $|\mathrm{in}\rangle$
and $|\mathrm{out}\rangle$ vacuum states and corresponding operators $a_{k}^{\mathrm{in},\mathrm{out}}$. The linear transformation between these two bases is given by the Bogoliubov coefficients $\alpha_{k},\beta_{k}$
through $\phi_{k}^{\mathrm{out}}(t)=\alpha_{k}(t)\phi_{k}^{\mathrm{in}}(t)+\beta^{*}_{k}\phi_{k}^{\mathrm{in*}}(t)$,
whereas the relation between vacuum states is given by the squeezed states,
$|0_{k},\mathrm{in}\rangle=|\alpha_{k}|^{-1}\exp(-(\beta_{k}/ \alpha^{*}_{k})a^{\mathrm{out}\dag}_{k}a^{\mathrm{out}\dag}_{-k}) |0_{k},\mathrm{out}\rangle$.
The alternative definition of the Bogoliubov coefficients can be
then given via the Klein-Gordon scalar product:
$\alpha_{k}=(\phi_{k}^{\mathrm{in}},\phi_{k}^{\mathrm{out}})$ and
$\beta_{k}^{*}=-(\phi_{k}^{\mathrm{in*}},\phi_{k}^{\mathrm{out}})$. The number of
excitations is now naturally defined as
\beq
\langle
0_{\mathrm{in}}|a^{\mathrm{out}\dag}_{k}a^{\mathrm{out}}_{k}|0_{\mathrm{in}}\rangle=|\beta_{k}|^{2}.
\eeq

The flat de Sitter evolution is defined as a condition that $H$ be a constant, which implies for the scale factor  $a(t)=H^{-1}\exp(Ht)$. Apparently, in this case $\epsilon\equiv 0$. The solution for the
Klein-Gordon equation is then given by $\phi_{k}^{\mathrm{in}}(t)=\sqrt{\pi/4H}e^{-\pi\nu/2}{\cal
H}_{i\nu}^{(1)}(k/[a(t)H])$ where ${\cal H}^{(1)}_{i\nu}(z)$ is a Hankel function of the first kind. Here
$\nu=\sqrt{(m/H)^{2}-(d/2)^{2}}$. Using the asymptotics of the Hankel function for large $|z|$ we can check that this solution indeed describes an $|\mathrm{in}\rangle$ oscillating state in conformal time $\eta=-\exp(-Ht)$, according to the picture of Ref.~\cite{BD}. The out solution has the following form:
\beq
\phi_{k}^{\mathrm{out}}(t)=\sqrt{\frac{\pi}{2H\sinh(\pi\nu)}}J_{i\nu}\left(\frac{k}{a(t)H}\right),
\eeq
where $J_{i\nu}(z)$ is a Bessel function. Now, the small-argument
asymptotics corresponds to the $|\mathrm{out}\rangle$ state. Using the
relation between the Hankel and Bessel functions,
\beq
e^{-\pi\nu}{\cal
H}^{(1)}_{i\nu}(z)=\frac{e^{\pi\nu}J_{i\nu}(z)-e^{-\pi\nu}J_{-i\nu}(z)}{\sinh(\pi\nu)}
\eeq
we obtain the {\it mode-independent} Bogoliubov coefficients
\beq
\alpha =\frac{1}{\sqrt{1-\exp(-2\pi\nu)}},\qquad \beta
=\alpha\exp(-\pi\nu).
\eeq

We consider now a regime where\footnote{Note that we use dimensionless notations. In physical units the dimensionless
combination is $mc^{2}/(\hbar H)$, and thus establishes a ratio between the Compton wavelength $\hbar/mc$ and the Hubble length $c/H$. Although $H$ is not a constant, $H\equiv H(t)$, in the present cosmological epoch it is $H = 74.2\pm 3.6$(km/s)/Mpc. The adiabaticity condition is thus justified sufficiently far from the "phase transition" point, in this case a Big Bang. }
\beq\label{delta}
\upsilon=H/m \ll 1
\eeq
which we identify as an adiabatic regime for reasons which will be clear immediately. In the regime of validity of (\ref{delta}), $\nu\approx 1/\upsilon$ and one obtains a density distribution of a pair production process
at late times ($\eta\rightarrow 0$) in the form of a thermal spectrum
\beq
n_{\mathrm{ex}}=1/[\exp(2\pi/\upsilon)-1].
\eeq
The probability of transition between asymptotic in and out vacua is given by the overlap, $P_{\mathrm{in}\rightarrow
\mathrm{out}}=\prod_{k}|\alpha_{k}|^{-2}$, and therefore in our case, the probability per each mode $k$ is given by
\beq
P_{in\rightarrow out}^{(k)}=1-\exp(-2\pi/\upsilon),
\eeq
which is nothing other than the LZMS transition probability~\cite{LZMS}. As we approach the Big Bang singularity the parameter $\upsilon$ starts to diverge and we expect that the adiabaticity conditions become violated.  We thus expect that the excitations are created at a much higher rate near the singularity, quite similar to what happens in adiabatic dynamics near QCPs.   Another possible source of adiabatic/non-adiabatic effects can be found by extending the simple de Sitter solution. In the initial stage of inflation, when the potential energy dominates the kinetic term, their ratio defines a so-called slow-roll parameter which is considered to be small; this parameter is strictly zero in the de Sitter case, while it is finite in more general metrics.  This slow-roll parameter can play the role of $\upsilon$, which defines the degree of non-adiabaticity in the system.

\subsection{Time Evolution in a Singular Space-time}

The idea we put forward in this section can be summarized as follows: suppose we have a quantum system which evolves on a curved space with a {\it time-dependent} metric. Suppose that this metric has singularities in some finite number of points. It is clear intuitively that the presence of these singularities must inevitably appear in the dynamics of the system. In some cases, as we will demonstrate below, nontrivial {\it geometry} can mimic nontrivial {\it dynamics}, similar to dynamics across a QCP.  A time-dependent metric of the background geometry plays the role of changing the external parameter and thus can induce non-adiabatic effects in quantum dynamics.

Our simple model here is inspired by recent interest in string theory literature on dynamics of quantum fields, mainly of string origin, in time-dependent geometries which contain some singularities.  Examples include time-dependent orbifolds, null-branes, pp-wave geometries, Big Crunch and Big Rip singularities (see Refs.~\cite{CE, LMS, N, TT}). The simplest illustrative example is the so-called Milne geometry described by the following metric in $1+1$ dimensions:
\beq
ds^{2}=-dt^{2}+\upsilon^{2}t^{2}dx^{2}.
\eeq
We consider a quantum scalar field propagating in this metric. Its
action is given by
\beq
S=\frac{1}{2}\int dx\int dt\,
|t\upsilon|\left[(\partial_{t}\phi)^{2}-\frac{(\partial_{x}\phi)^{2}}{t^{2}\upsilon^{2}}-m^{2}\phi^{2}\right]\,,
\eeq
where we assume that the evolution starts at $t=-\infty$ and ends at
$t=+\infty$. We introduced the scale factor $\upsilon$ to emphasize its role as the adiabatic parameter. In principle it can be set to unity by appropriately rescaling space-time units. The quantum Hamiltonian corresponding to this action is therefore
\beq
H=\frac{1}{2|t\upsilon|}\int dx
(\Pi^{2}+(\partial_{x}\phi)^{2})+\frac{m^{2}|t\upsilon|}{2}\int
dx\phi^{2}
\eeq
where $\Pi$ is the canonically conjugate momentum.

Apparently, this type of a system can be given a simple condensed-matter interpretation: a massive scalar field describes a large variety of one-dimensional phenomena. For example, a Luttinger liquid described by the interaction parameter $K\equiv |\upsilon t|$ perturbed by some relevant perturbation (which in the strong-coupling regime of
corresponding RG can be approximated by a quadratic massive term) could provide a realization of one of the physical models. In our case, both the interaction parameter $K$ and the strength of the perturbation depend explicitly on time. From the flat-space point of view it is therefore a non-equilibrium model.

The equations of motion for the model can be put into the form of time-dependent oscillators describing different momentum modes:
\beq
\ddot{\varphi}_{k}+\Omega_{k}(t)\varphi_{k}=0,
\eeq
where $\varphi=\sqrt{t}\phi$ and $\Omega_{k}(t)=m^{2}+[(k/\upsilon)^{2}+1/4]/t^{2}$.} The solution can
be given in terms of the Bessel functions of the order $\nu=ik/\upsilon$ of the argument $mt$.

We note that the situation is rather similar to the previous subsection, except that in the present case the quantum evolution goes across the singularity. The presence of the singularity implies that the wave functions of the system for $t<0$ and for $t>0$ must be properly defined and related to each other at the singularity. Several different ways to do so have been suggested in the literature: geometric (by going into covering space), operator-analytic (by properly regularizing the singularity), dimensional, etc. Without discussing this issue in detail (physically-relevant quantities should not depend on the regularization prescription anyway) let us simply state the Bogoliubov coefficients, which can be defined again by the matching of in and out states:
\beq
\alpha_{k}=-\frac{\cos(A-i\pi k/\upsilon) 
}{\sinh(\pi k/\upsilon)},\qquad\beta_{k}=i\frac{\cos(A)}{\sinh(\pi
k/\upsilon)}
\eeq
where the pure phase factor $A$ plays the role of the reflection
coefficient and can be defined from the mode-matching condition at
the singularity, $A=\mathrm{arg}[J_{ik/\upsilon}(mt)/J_{-ik/\upsilon}(-mt)]$.

The number of particles produced is given by $|\beta_{k}|^{2}$ and can be characterized as a thermal distribution with some mass-dependent temperature $T(m)$. For large $m$, $T(m)$ approaches a constant independent of $k$ and therefore $n_{\mathrm{ex}}\sim e^{-\pi m}$ whereas for small $m$ it is independent of $m$ and $n_{\mathrm{ex}}\sim
e^{-2\pi k/\upsilon}$. Therefore in the latter case the total number of particles $N=\sum_{k}n_{\mathrm{ex}}(k)$ scales as $|\upsilon|$. Correspondingly, the in-out transition probability can be put into the LZMS form with $k$-dependent effective velocity. For large $m$, $P\sim 1-e^{-\pi m}$ independent of $k$, whereas for small $m$,
$P\sim 1-e^{-2\pi k/\upsilon}$ independent of $m$.

We therefore conclude that evolution on a singular time-dependent manifold is to some extent equivalent to quantum non-equilibrium dynamical systems evolving across a phase transition. In some cases non-equilibrium dynamics of quantum system with time-dependent parameters can be modeled by the evolution of a quantum system with time-independent parameters on a time-dependent background geometry. This geometry can have singularities which then correspond to QCPs in the quantum system.

\section{Summary and Outlook}

In this chapter we gave an overview of different connections between certain universal equilibrium and non-equilibrium properties of continuous quantum phase transitions. It is well known that such transitions are typically characterized by singularities in thermodynamic quantities, in particular, in various susceptibilities. We discussed here that these singularities lead to the universal nonlinear dynamical response of these observables to
various dynamical processes near a QCP.

In particular, we analyzed in detail two possibilities: a sudden spatially uniform quench starting at a QCP; and slow passage through a QCP. In the latter case one can also start (end) right at the QCP. In low dimensions such quantities as heat (excess energy) generated in the system or density of generated quasi-particles become universal non-analytic functions of the adiabatic parameter, which is quench amplitude for sudden quenches, quench rate for slow linear quenches, etc. In particular, if the quench $\lambda(t)\sim \upsilon t^r$ then the quasi-particle density scales as $n_{\rm ex}\sim |\upsilon|^{d\nu/(z\nu r+1)}$.  A similar scaling is valid for heat and for other thermodynamic quantities. These scalings are expected to be generically valid as long as the corresponding exponents of $\upsilon$ remain smaller than two. Otherwise, the leading asymptotic behavior becomes quadratic, consistent with standard perturbation theory: $n_{\rm ex}\sim \upsilon^{2}$. Crossover from non-perturbative to perturbative scaling happens precisely at the point where the scaling dimension of the corresponding generalized adiabatic susceptibility $\chi_{2r+2}(\lambda)$ (see Eq.~(\ref{chim}))  becomes zero, i.e., when it becomes finite at the QCP. For sudden quenches the relevant susceptibility determining $n_{\rm ex}$: $\chi_2(\lambda)$ is nothing but the fidelity susceptibility; the latter became recently an interesting new measure of quantum criticality,
independent of the choice of the observable. For slow quenches characterized by the exponent $r$ the relevant susceptibilities determining the scaling of the quasiparticle density and the probability of exciting the system $\chi_{2r+2}(\lambda)$ is thus a direct generalization of the fidelity susceptibility. In particular,  the scaling for linear quenches $n_{\rm ex}\sim |\upsilon|^{d\nu/(z\nu+1)}$, which can be also explained by the Kibble-Zurek
arguments,~\cite{zurek_adiabatic,kz1, kz2}, is associated with the scaling dimension of the susceptibility $\chi_4(\lambda)$. It would be very interesting to analyze situations in the future in which many-body excitations do not directly correspond to a fixed number of quasi-particles. In those situations one still expects that the scaling for the density of defects will deviate from Eq.~(\ref{nex1}).  A possible candidate for this scenario is the non-adiabatic regime, where the density of created excitations diverges in the thermodynamic limit~\cite{pg_np}.

We illustrated how scaling results emerge from adiabatic perturbation theory. This theory has a purely geometric
interpretation since time can be dropped completely from the analysis (it only implicitly enters through the rate of change of the coupling). We discussed a somewhat simplistic situation where only one coupling changes in time.  One can imagine more general scenarios where several coupling constants change simultaneously. Then the expression for the transition amplitude within adiabatic perturbation theory represents a contour integral in the parameter
space. In this case the expectation values of dynamical quantities will be related to the quantum geometric tensors in the Hilbert space of a system. These objects define a structure of the (complex) Riemannian metrics on the space of parameters of a system. Evaluated in the ground state, the real part of this metrics is related to the (generalized) fidelity susceptibilities, whereas its imaginary part is related to the adiabatic Berry curvature. In adiabatic
perturbation theory the dynamical phase should then be modified by inclusion of the geometric Berry phase. The Berry curvature can diverge close to QCPs. It is then clear that universal features of evolution close to the phase transition advocated here should be corrected by specifying the path of the quantum evolution in the parameter space. In particular, one can expect additional corrections to the scaling laws coming from the singularities of the
Berry phase~\cite{tomka}. We expect an interesting interplay between dynamical and geometrical effects in the scaling dependence of quantities in the linear-quench regime. Such a possibility to reduce critical dynamics
to statics quantities, like quantum geometric tensors, looks very intriguing and perhaps requires a closer look beyond adiabatic perturbation theory.

We also showed that the universal dynamical response can be strongly affected by initial thermal fluctuations and that
quasi-particle statistics changes the scaling laws. We discussed somewhat simplistic situations where quasi-particles are either non-interacting bosons or fermions. In general, critical dynamics at finite temperatures in the quantum critical region remains an open problem and it is clear that extra input from equilibrium properties to the dynamical response is needed, as compared to the zero temperature case.

In the last section we gave a brief outlook of connections of critical dynamics with other areas of physics like cosmology and string theory. Dynamics near QCPs is qualitatively similar to dynamics near various space-time
singularities. In some cases non-equilibrium dynamics of a quantum system with time-dependent parameters can be modeled by the evolution of a quantum system with time-independent parameters. However, the background geometry of space-time is explicitly time-dependent. Quantum dynamics close to singularities of this dynamical geometry may correspond to crossing the phase transition in the static geometry.

There are many other open questions remaining. The main purpose of this chapter was to shed light on some non-equilibrium universal aspects of dynamics near critical points beyond perturbation theory and show
their close connections to static equilibrium properties. We hope that this chapter will partly stimulate further research in this exciting new area.

{\em Acknowledgements.} We acknowledge discussions with R. Barankov and C. De Grandi. A.P. was supported through the AFOSR YIP, the National Science Foundation (DMR-0907039), and the Sloan Foundation. V.G. was supported by the Swiss NSF. A.P. also acknowledges hospitality of the Aspen Center for Physics where part of this work was done.


\begin{thebibliography}{10}

\bibitem{sachdev_book}
S. Sachdev, {\em Quantum Phase Transitions} (Cambridge University Press,
  Cambridge, 1999).

\bibitem{sondhi1997rmp}
S.~L. Sondhi, S.~M. Girvin, J.~P. Carini, and D. Shahar, Rev. Mod. Phys. {\bf
  69},  315  (1997).

\bibitem{vojta2003progphys}
M. Vojta, Rep. Prog. Phys. {\bf 66},  2069  (2003).

\bibitem{greiner2002nature}
M. Greiner, O. Mandel, T. Esslinger, T.~W. Hansch, and I. Bloch, Nature {\bf
  415},  39  (2002).

\bibitem{fisher1989}
M.~P.~A. Fisher, P.~B. Weichman, G. Grinstein, and D.~S. Fisher, Phys. Rev. B
  {\bf 40},  546  (1989).

\bibitem{cardy_quench1}
P. Calabrese and J. Cardy, Phys. Rev. Lett. {\bf 96},  136801  (2006).

\bibitem{cardy_quench2}
P. Calabrese and J. Cardy, J. Stat. Mech: Th. and Exp. {\bf P06008},    (2007).

\bibitem{ap_adiabatic}
A. Polkovnikov, Phys. Rev. B {\bf 72},  R161201  (2005).

\bibitem{zurek_adiabatic}
W.~H. Zurek, U. Dorner, and P. Zoller, Phys. Rev. Lett. {\bf 95},  105701
  (2005).

\bibitem{reimann}
P. Reimann, Phys. Rev. Lett. {\bf 101},  190403  (2008).

\bibitem{olshanii_nature}
M. Rigol, V. Dunjko, and M. Olshanii, Nature {\bf 452},  854  (2008).

\bibitem{pg_np}
A. Polkovnikov and V. Gritsev, Nat. Phys. {\bf 4},  477  (2008).

\bibitem{claudia_kolk} C. De\;\;Grandi and A. Polkovnikov, ``Adiabatic perturbation theory: from Landau-Zener problem to quenching through a quantum critical point'' in ``Quantum Quenching, Annealing and Computation'', Eds. A. Das, A. Chandra and B. K. Chakrabarti, Lect. Notes in Phys., vol. 802 (Springer, Heidelberg 2010), arXiv:0910.2236.


\bibitem{kz1}
T.~W.~B Kibble, J. Phys. A {\bf 9},  1387  (1976).

\bibitem{kz2}
W.~H. Zurek, Phys. Rep. {\bf 276},  177  (1996).

\bibitem{venuti}
L.~C. Venuti and P. Zanardi, Phys. Rev. Lett. {\bf 99},  095701  (2007).

\bibitem{gu}
S.-J. Gu and H.-Q. Lin, Europhys. Lett. {\bf 87},  10003  (2009).


\bibitem{alet} D. Schwandt, F. Alet, and S. Capponi, Phys. Rev. Lett. {\bf 103}, 170501 (2009).

\bibitem{quench_qcp} C. De\:\:Grandi, V. Gritsev, and A. Polkovnikov Phys. Rev. B {\bf 81}, 012303 (2010); {\em ibid} arXiv:0910.0876.


\bibitem{diag_entropy} A. Polkovnikov, arXiv:0806.2862.

\bibitem{ap_heat}
A. Polkovnikov, Phys. Rev. Lett. {\bf 101},  220402  (2008).

\bibitem{ortiz_2008}
G. Rigolin, G. Ortiz, and V.~H. Ponce, Phys. Rev. A {\bf 78},  052508  (2008).

\bibitem{sengupta2008}
D. Sen, K. Sengupta, and S. Mondal, Phys. Rev. Lett. {\bf 101},  016806
  (2008).

\bibitem{barankov2008}
R. Barankov and A. Polkovnikov, Phys. Rev. Lett. {\bf 101},  076801  (2008).

\bibitem{kitaev}
A. Kitaev, Ann. Phys. {\bf 321},  2  (2006).

\bibitem{antonio_rmp} A.~H.~Castro Neto, F.~Guinea, N.~M.~R.~Peres, K.~S.~Novoselov, and A.~K.~Geim,
Rev. Mod. Phys. {\bf 81}, 109 (2009).

\bibitem{dziarmaga}
J. Dziarmaga, Phys. Rev. Lett. {\bf 95},  245701  (2005).

\bibitem{ortiz2008}
G. Rigolin, G. Ortiz, and V.~H. Ponce, Phys. Rev. A {\bf 78},  052508  (2008).

\bibitem{shankar}
R. Shankar, {\em Principles of Quantum Mechanics} (Springer, New York, ADDRESS,
  1994).

\bibitem{tomka}
M. Tomka, V. Gritsev, A. Polkovnikov, Geometric effects in quantum
  nonequilibrium dynamics, in preparation.

\bibitem{kollar_09} M.~Eckstein and M.~Kollar, arXiv:0911.1282.

\bibitem{silva2009}
D. Rossini, A. Silva, G. Mussardo, and G.~E. Santoro, Phys. Rev. Lett. {\bf
  102},  127204  (2009).

\bibitem{cardy2009}
S. Sotiriadis, P. Calabrese, and J. Cardy, Europhys. Lett. {\bf 87},  20002
  (2009).

\bibitem{bloch_rmp}
I. Bloch, J. Dalibard, and W. Zwerger, Reviews of Modern Physics {\bf 80},  885
   (2008).

\bibitem{LL5}
L.~D. Landau and E.~M. Lifshitz, {\em Statistical Physics, Third Edition, Part
  1: Volume 5} (Butterworth-Heinemann, ADDRESS, 1980).

\bibitem{Lukyanov97}
S. Lukyanov, Mod. Phys. Lett. {\bf A12},  2543  (1997).

\bibitem{Linde}
A. Linde, Lect. Notes Phys. {\bf 738},  1  (2008).


\bibitem{BD}
T.S. Bunch and P.~C.~W. Davies, Proc. R. Soc. Lond. Ser. A {\bf 360},  117
  (1978).


\bibitem{LZMS} L. Landau, Phys. Z. Sowjetunion {\bf 1},  88  (1932); C. Zener, Proc. R. Soc. London, Ser. A {\bf 137},  696  (1932); E. Majorana, Nuovo Cimento {\bf 9},  43  (1932); E.~C.~G. St\"{u}ckelberg, Helv. Phys. Acta {\bf 5},  369  (1932).



\bibitem{CE}
B. Craps and O. Evnin, JHEP {\bf 0804},  021  (2008).

\bibitem{LMS}
H. Liu, G. Moore, and N. Seiberg, JHEP {\bf 0206},  045  (2002).

\bibitem{N}
N.~A. Nekrasov, Surveys High Energ.Phys. {\bf 17},  115  (2002).

\bibitem{TT}
A.~J. Tolley and N. Turok, Phys. Rev. D {\bf 66},  106005  (2002).

\end{thebibliography}
\end{document}